\documentclass[11pt]{article}
\usepackage{amssymb,amsmath}

\newcommand{\ft}[2]{{\textstyle\frac{#1}{#2}}}

\newcommand{\pa}{\partial}
\newcommand{\be}{\begin{equation}}
\newcommand{\ee}{\end{equation}}
\newcommand{\bea}{\begin{eqnarray}}
\newcommand{\eea}{\end{eqnarray}}

\newcommand{\T}{\mbox{Tr}}
\newcommand{\bD}{\bar{D}}

\newcommand{\cN}{{\cal N}}

\newcommand{\cO}{{\cal O}}

\newcommand{\hp}{\hat\partial}

\renewcommand{\a}\alpha

\newcommand{\da}{\dot\alpha}
\renewcommand{\b}{\beta}
\newcommand{\g}{\gamma}
\newcommand{\db}{\dot\beta}

\newcommand{\q}{\theta}
\newcommand{\bq}{\bar\q}
\newcommand{\bz}{\bar{z}}
\newcommand{\ep}{\epsilon}

\newcommand{\bt}[1]{{\bar t}}
\newcommand{\ts}{\textstyle}

\newcommand{\half}{{\ts \frac{1}{2}}}

\input epsf
\usepackage{amsmath,amsfonts,epsfig,color,latexsym}
\usepackage{amssymb}

\begin{document}

\thispagestyle{empty}

\null\vskip-12pt \hfill  LAPTH-1109/05 \\
\vskip0.2truecm
\begin{center}
\vskip 0.2truecm {\Large\bf
\huge{On twist-two operators in ${\cal N}=4$ SYM }
}\\
\vskip 1truecm
{\bf\large J. Henn, C. Jarczak and E. Sokatchev \\
}

\vskip 0.4truecm

\vskip .2truecm {\it Laboratoire d'Annecy-le-Vieux de
Physique Th\'{e}orique  LAPTH,\\
B.P. 110,  F-74941 Annecy-le-Vieux, France\footnote{UMR 5108
associ\'{e}e {\`a}
 l'Universit\'{e} de Savoie} } \\
\end{center}

\vskip 2truecm \Large
\centerline{\bf Abstract} \vskip 1truecm\normalsize We propose a mechanism for calculating anomalous dimensions of higher-spin twist-two operators in ${\cal N}=4$ SYM. We consider the ratio of the two-point functions of the operators and of their superconformal descendants or, alternatively, of the three-point functions of the operators and of the descendants with two protected half-BPS operators. These ratios are proportional to the anomalous dimension and can be evaluated at $n-1$ loop in order to determine the anomalous dimension at $n$ loops. We illustrate the method by reproducing the well-known one-loop result by doing only tree-level calculations.\\
We work out the complete form of the first-generation descendants of the twist-two operators and the scalar sector of the second-generation descendants.

\newpage
\setcounter{page}{1}\setcounter{footnote}{0}

\section{Introduction}

The AdS/CFT correspondence triggered a renewed interest in the spectrum of anomalous dimensions of gauge invariant operators with higher spin. In particular, the higher-spin symmetry enhancement at small string radius \cite{HS} together with results for the ``first Regge trajectory" \cite{Sezgin}
 have permitted the exact matching of the full string spectrum with the operator spectrum of free ${\cN=4}$ super-Yang-Mills (SYM) theory \cite{BBMS}. At finite string radius most of the string states become massive and their CFT  twist-two higher-spin operator counterparts acquire anomalous dimension. The present paper is a modest attempt to understand the mechanisms for creation of this anomalous dimension.

Perturbative results for the anomalous dimension of twist-two operators have been available for some time. The one-loop values have been obtained through direct calculations \cite{Lipatov:1loop} and by OPE methods \cite{Dolan:1loop}. The two-loop values have been found again by a direct calculation in \cite{Kotikov:2loop} and by OPE analysis \cite{Dolan:2loop} of the four-point function of stress-tensor multiplets calculated in \cite{4pt2loop}. More recently, a conjecture about the three-loop values was made in \cite{Kotikov:3loop}, based on extensive QCD calculations \cite{Vogt:2004mw}. This latter result was in excellent agreement with the predictions of the dilatation operator approach \cite{dilop}. It was confirmed by a direct calculation in the simplest case of spin 0 in \cite{EJS}. A very interesting possible explanation of this three-loop spectrum, based on integrability and the factorization of the S matrix \cite{Arutyunov:2004vx}, was proposed in \cite{Staudacher:200
 4tk}.

The twist-two operators in ${\cN=4}$ SYM have rather special
properties if regarded as unitary irreducible representations (UIRs)
of the superconformal group $PSU(2,2/4)$. In the free theory
(vanishing coupling $g=0$) they are situated exactly at the
unitarity bound \cite{DP1} and realize reducible ``semishort"
representations. The irreducibility conditions take the form of a
``superconservation law" (schematically) $D^\a J_{\a\ldots}=0$,
where  $J$ is the twist-two operator (superfield) of spin $j$ and
canonical dimension $d=j+2$, and  $D$ is a spinor superspace
derivative. As soon as the interaction is switched on, they
``swallow" a few other representations of lower spin and higher
naive dimension to form together one ``long" irreducible multiplet
with anomalous dimension. These other representations now appear in
the right-hand side of the superconservation law as two generations
of superconformal descendants of the twist-two multiplet,
(schematically) $D^\a J_{\a\ldots}=gM_{\ldots}$ and $\bar
D^{\da}M_{\da\ldots}=gN_{\ldots}$. Such a mechanism of
disintegration of long multiplets into a set of semishort and short
multiplets was presented in \cite{Dolan:semishort} (see also
\cite{Dobrev:2004tk}), giving an alternative interpretation of the
semishort operators found in \cite{SemiShort}. Its applications in
the context of the AdS/CFT correspondence were studied in
\cite{Bianchi:2003wx,Bianchi:2004xi} under the name {\it La Grande
Bouffe}. In a recent paper \cite{BHR} some details of the mechanism
of decomposition of long twist-two multiplets into submultiplets
were presented, and in particular, the creation of two generations
of descendants described above.

In this paper we propose a way of computing the anomalous dimensions
of twist-two operators which exploits {\it La Grande Bouffe}
mechanism. The basic idea goes back to a paper by Anselmi
\cite{Anselmi} who used the non-conservation of the Konishi current
to compute the one-loop anomalous dimension of the Konishi multiplet
(and also of a few of the higher-spin twist-two operators). He
noticed that in a conformal theory the two-point function of the
divergence of a (non-conserved) current is proportional to the
anomalous dimension $\g(g^2) = \g_1 g^2 + \g_2 g^4 + \g_3 g^6 +
\ldots$. This implies that the ratio of the two-point functions of
the divergence of the current and of the current itself, evaluated
at {\it tree level}, determines the one-loop value $\g_1$ without
doing any Feynman integrals. In other words, by comparing two-point
functions of primary operators and of their conformal descendants,
we can effectively gain one order in perturbation theory. This idea
was successfully used in \cite{Eden:2003sj} to compute $\g_2$ for
some operators of the BMN series by doing only one-loop
calculations. Further, the same method made it possible to obtain
$\g_3$ for the Konishi operator (the spin zero twist-two operator)
and for one of the BMN operators by means of a two-loop calculation
\cite{EJS}.

Here we adapt the method to the case of higher-spin twist-two operators. We propose two scenarios. The first one is a close analog of the two-point mechanism above. We consider the two-point functions $\langle JJ \rangle$ and $\langle M\bar M \rangle$ of the primary twist-two operator $J$ and of its first-generation descendant $M$. Their ratio is again proportional to the anomalous dimension,
\begin{equation}\label{0}
  g^2 \frac{\langle M\bar M \rangle}{\langle JJ \rangle} \propto \g_1 g^2 +  \ldots\ .
\end{equation}
Thus, evaluating the two-point functions at tree level, we obtain the one-loop value $\g_1$. In order to do this, we need the explicit realization of the operators $J$ and $M$ in terms of the elementary fields of the $\cN=4$ SYM theory. The form of $J$ was found in \cite{BHR}, but we present an alternative derivation. We also work out the complete form of the descendant $M$ by applying to $J$ on-shell supersymmetry transformations in terms of component fields.

Our second scenario uses three-point functions of the primary $J$ and of its second-generation descendant $N$ with two protected half-BPS operators $\cO$. Once again, the ratio
\begin{equation}\label{0'}
  g^2 \frac{\langle \cO\bar \cO N \rangle}{\langle \cO\bar \cO J \rangle} \propto \g_1 g^2 +  \ldots
\end{equation}
allows us to determine $\g_1$ through tree-level calculations. We show that this second approach is considerably simpler than the first. The reason is the linear insertion of the primary $J$ and especially of its descendant $N$ into the two-point function of protected operators $\cO$. The resulting combinatorial sums are much simpler than in the case of the two-point function $\langle M\bar M \rangle$, quadratic in the rather complicated descendant $M$. For carrying out the tree-level calculation we need only the scalar sector in the descendant $N$. The complete expression for $N$ could be worked out via a straightforward, although tedious calculation, most efficiently done in the component field framework.

Our goal in this paper is mainly methodological. We explore the possibility of applying the superdescendant mechanism to the computation of anomalous dimensions of higher-spin operators and illustrate it by deriving the well-known one-loop result. The next step would be to obtain the two-loop anomalous dimension by means of a one-loop calculation of two- or three-point functions involving descendants. This will require a careful study of the possible quantum anomalies, i.e. of the operator mixing problem at the level of the descendants. A more ambitious goal would be to confirm the conjecture of \cite{Kotikov:3loop} about the three-loop value and, if possible, get some insight into the origin of the anomalous dimension of higher spins. We also hope that our method will prove useful in {\it La Grande Bouffe} scenario.

\newpage

\section{General superconformal considerations}

\subsection{Semishortness conditions and twist-two operators}

The classification of the UIRs of the superconformal group $PSU(2,2/4)$ has been carried out in \cite{DP1}. In particular, it has been found that there exists a continuous series of representations characterized by their conformal dimension $d$, Lorentz spins $(j_1,j_2)$ and $SU(4)$ Dynkin labels $[a_1,a_2,a_3]$.  For them unitarity requires that
\begin{equation}\label{1}
  d \geq 2 + (j_1+j_2) + (a_1+a_2+a_3)\,.
\end{equation}
When the unitarity bound is saturated, these representations become reducible
and one can impose irreducibility constraints.

The subject of this paper are the twist-two operators of spin $j$. They belong to the continuous series (\ref{1}) with Dynkin labels $[0,0,0]$ ($SU(4)$ singlets) and $j_1=j_2=j/2$. In $\cN=4$ superspace they are described
by real superfields $J_{\a_1\cdots\a_j\ \da_1\cdots\da_j}(x, \q^A,
\bq_A)$ ($A=1,2,3,4$) carrying $j$ totally symmetrized dotted and undotted
two-component Lorentz spinor indices.\footnote{Equivalently, one can represent $J$ as a symmetric traceless tensor of rank $j$, $J^{\mu_1\cdots\mu_j}$. } If the bound (\ref{1}) is
saturated, $d_0=2+j$ (this takes place in the free field theory), the irreducibility conditions on
$J$ read \cite{DP1,DP2,Park:1999pd}
\begin{equation}\label{2}
  D^\a_A J_{\a\a_2\cdots\a_j\ \da_1\cdots\da_j} = \bD^{\da\, A} J_{\a_1\cdots\a_j\ \da\da_2\cdots\da_j} = 0\,.
\end{equation}
Here
\begin{equation}\label{1008}
  D^\a_A = - \frac{\pa}{\pa \q^A_\a} - i\bq_{A\,\da} \pa^{\a\da}\,, \qquad \bD^{\da\, A} =   \frac{\pa}{\pa \bq_{A\,\da}} +  i\q^{A}_\a \pa^{\a\da}
\end{equation}
are the supersymmetric spinor derivatives satisfying the supersymmetry algebra
\begin{equation}\label{2'}
  \{D^\a_A,\bar{D}^{\da\, B}\} = -2i \delta^B_A\,\pa^{\a\da}\,, \qquad \pa^{\a\da} = \sigma_\mu^{\a\da}\pa^\mu\,.
\end{equation}
An immediate corollary of this is the ``current conservation"
\begin{equation}\label{3}
  \pa^{\a\da}J_{\a\a_2\cdots\a_j\ \da\da_2\cdots\da_j} = 0\,.
\end{equation}
However, in an interacting theory such ``currents" are no longer
conserved, and eqs.\;(\ref{2}) have non-vanishing right-hand sides \cite{BHR}:
\begin{eqnarray}\label{4}
  D^\a_{A} J_{{\a\a_2\cdots\a_j\ \da_1\da_2\cdots\da_j}} = g M_{A\ {\a_2\cdots\a_j\ \da_1\da_2\cdots\da_j}}\,,
  \qquad \bar{D}^{\da\, A} J_{\da \ldots} = g
  \bar{M}^A_{\ldots}\ ,
\end{eqnarray}
where $g$ is the coupling constant. We can say that eq.\,(\ref{4}) defines a first-generation superconformal ``descendant" with spins $((j-1)/2,\ j/2)$ and Dynkin labels $[1,0,0]$. Another first-generation descendant with $((j-1)/2,\ (j-1)/2)$ and $[0,0,0]$ is obtained from the violation of the conservation condition (\ref{3}),
\begin{equation}\label{4'}
  \pa^{\a\da} J_{\a\a_2\cdots\a_j\ \da\da_2\cdots\da_j} = gK_{\a_2\cdots\a_j\ \da_2\cdots\da_j}\,.
\end{equation}
One can go on and produce a second-generation descendant with $((j-1)/2,\ (j-1)/2)$ and $[1,0,1]$:\footnote{In the free theory ($g=0$) $M$, $K$ and $N$ decouple from their ``parent" $J$ and become independent multiplets. In particular, $M$ and $N$ saturate the corresponding unitarity bounds and are thus subject to their own irreducibility conditions (one of them is (\ref{4001}) at $g=0$). This explains why $N^A_B$ has to be traceless, i.e. in the $SU(4)$ representation $[1,0,1]$: It has canonical dimension $d_0=j+3$ in accord with (\ref{1}), $d_0 = 2+ ((j-1)/2+(j-1)/2) + (1+0+1)$. }
\begin{equation}\label{4001}
  \bar{D}^{\da\, A}{M}_{B\ \da\ldots} - \frac{1}{4}\delta^A_B \bar{D}^{\da\, C}{M}_{C\ \da\ldots} = -D^\a_{B}\bar{M}^A_{\a\ldots} + \frac{1}{4}\delta^A_B D^\a_{C}\bar{M}^C_{\a\ldots} = g N^A_{B\ \ldots}\,,
\end{equation}
or equivalently,
\begin{equation}\label{4002}
  [\bar{D}^{\da\, A},D^\a_{B}]J_{\a\da \ldots}  - \frac{1}{4}\delta^A_B [\bar{D}^{\da\, C},D^\a_{C}]J_{\a\da \ldots} = 2g^2N^A_{B\ \ldots}\ .
\end{equation}
At the same time, the superconformal primary operator $J$ and its descendants $M$, $K$ and $N$ ``drift
away" from the unitarity bound by acquiring an anomalous dimension $\gamma(g^2)$:
\begin{equation}\label{5}
  d_J=j+2 + \gamma \,, \qquad d_{M}=j+5/2 + \gamma \,, \qquad d_{K}= d_{N}=j+3 + \gamma \,.
\end{equation}

We end this subsection by remarking that, from an abstract point of view, the spin $j$ can take even or odd values. However, the explicit realization of the semishort operators $J$ with the defining property (\ref{2}) in terms of the elementary fields of the $\cN=4$ SYM theory is only possible for $j$ even (see Section \ref{TTO}).

\subsection{BPS shortness and protected operators}

In this paper, besides the twist-two operator $J$ and its descendants, we also use another operator $\cO$ belonging to the discrete series \cite{DP1} of the so-called ``BPS short" UIRs for which
\begin{equation}\label{1001}
  j_1=j_2=0\,, \qquad d = a_1+a_2+a_3\,.
\end{equation}
In particular, when $a_1=a_3=0$, $a_2=k$ we have 1/2 BPS short
multiplets of fixed (quantized) conformal weight $d=k$ which are
annihilated by half of the supercharges (or spinor derivatives),
\begin{equation}\label{1002}
  D_3\cO = D_4\cO = \bD^1\cO= \bD^2\cO= 0\,.
\end{equation}
In fact, these are conditions for ``Grassmann analyticity" \cite{GIO}, a consequence of the field equations of the $\cN=4$ SYM theory, as explained below.

In $\cN=4$ superspace the free on-shell SYM multiplet is described by the field-strength superfield $W^{AB}(x,\q^C,\bq_C) = -W^{BA} = \phi^{AB}(x)+\mbox{$\q$ terms}$. It belongs to the $\mathbf{6}$ of $SU(4)$ and satisfies the free on-shell constraints (field equations)
\begin{equation}\label{501}
  D_{\a\, \{ C} W^{A\}B} = 0\,, \qquad \bar D^{\da\, (C} W^{A)B}=0\,,
\end{equation}
where ${}_{\{C}{}^{A\}}$ denotes the traceless part and $(CA)$ means symmetrization. In addition, this superfield is real, $\overline{W^{AB}} = W_{AB} = \frac{1}{2} \epsilon_{ABCD} W^{CD}$.

The constraints (\ref{501}) can be rewritten in the form of Grassmann analyticity. Take, for instance, the highest-weight $SU(4)$ projection $W^{12}$, for which (\ref{501})  implies the constraints
\begin{equation}\label{502}
  D_3 W^{12} = D_4 W^{12} = \bar D^1  W^{12} = \bar D^2  W^{12} = 0\,.
\end{equation}
In fact, they are equivalent to the full set (\ref{501}). Indeed, the highest-weight state $W^{12}$ is annihilated by all the raising operators of $SU(4)$, $T_+ W^{12} = 0$ ($T_+ \equiv T^A_B$, $A<B$), while the other projections of $W^{AB}$ can be obtained with the help of the lowering operators $T_-$ ($T_- \equiv T^A_B$, $A>B$). In the same way, starting with (\ref{502}) and applying $T_-$, we obtain all the other projections of (\ref{501}).

The advantage of the analytic form (\ref{502}) of the constraints is that they can be solved in an appropriate superspace basis,
\begin{equation}\label{503}
     X^{\mu} = x^\mu + i(\theta^{1}\sigma^\mu\bar\theta_1 + \theta^{2}\sigma^\mu\bar\theta_2) - i(\theta^{3}\sigma^\mu\bar\theta_3 + \theta^{4}\sigma^\mu\bar\theta_4)\,,
\end{equation}
in which the spinor derivatives (\ref{1008}) appearing in (\ref{502}) become ``short", i.e. partial derivatives with respect to $\q^{3,4}$ and $\bq_{1,2}$. As a consequence, $W^{12}$ depends only on half of the odd coordinates (hence the term ``Grassmann analyticity"):
\begin{eqnarray}
  W^{12}(X, \q^{1,2},\bq_{3,4}) &=& \phi^{12}(X) -i\sqrt{2}\ \q^{\a\,[1}\lambda_\a^{2]}(X) +i\sqrt{2}\ \bq_{\da\, [3}\bar\lambda^{\da}_{4]}(X)\nonumber\\
  &+&   i/\sqrt{2}\ \q^{\a\,[1}\q^{\b\, 2]} F_{\a\b}(X)  -  i/\sqrt{2}\ \bq_{\da\,[3}\bq_{\db\, 4]} \bar F^{\da\db}(X)\nonumber\\
  &+&  \mbox{derivative terms}\ , \label{504}
\end{eqnarray}
where $\q^{\a\,[1}\lambda_\a^{2]} = \q^{\a\,1}\lambda_\a^{2} -
\q^{\a\,2}\lambda_\a^{1}$, etc. Here we can see $SU(4)$ projections
of the component fields of the $\cN=4$ SYM multiplet, the six
scalars $\phi^{AB}(x)$, the four fermions $\lambda^A_\a(x)$,
$\bar\lambda^{\da}_{A}(x)$ and the gluon field-strength $F_{\a\b}(x)$,
$\bar F_{\da\db}(x)$.

In addition, this multiplet is on shell, i.e. the fields in (\ref{504}) satisfy the corresponding equations of motion. The reason for this is  the $SU(4)$ highest-weight constraint $T_+ W^{12} = 0$, in which the generators $T_+$ now involve terms of the type $\q\sigma^\mu\bq\pa_\mu$, due to the change of variables (\ref{503}) (for more details see \cite{Zupnik}). Note that in the interacting theory the spinor derivatives in (\ref{502}) are replaced by gauge covariant ones, so the Grassmann analyticity of $W^{12}$ is not a manifest property of the interacting SYM multiplet.

Grassmann analyticity can be maintained manifest, even in the interacting theory, if we consider gauge invariant composite operators made out of $W^{12}$ \cite{HWest}. The simplest example is the  operator
\begin{equation}\label{505'}
  \cO=\T(W^{12}W^{12})\,.
\end{equation}
It satisfies the ``shortness" (irreducibility) conditions (\ref{1002}) as an obvious corollary of  the on-shell constraints (\ref{502}) on $W^{12}$. This operator is the  highest-weight $SU(4)$ projection of the so-called ``stress-tensor multiplet" of $\cN=4$ SYM,
\begin{equation}\label{505}
  (\cO_{\mathbf{20'}})^{ABCD} = \half\T\left(W^{AB}W^{CD} + W^{AD}W^{CB}  \right)\,,
\end{equation}
which belongs to the  $\mathbf{20'}$ of $SU(4)$ (Dynkin labels ${[0,2,0]}$). In other words, it is an operator of the half-BPS type (\ref{1001}), (\ref{1002}) with fixed conformal dimension $d=2$,  ``protected" from quantum corrections.

\section{Method for computing the anomalous dimension of twist-two operators}

In quantum field theory the anomalous dimension $\gamma$ of an
operator is usually associated with divergences in the Feynman
graphs for, e.g., the two-point function of the operator. In the
special case of (super)conformal operators near the unitarity bound
like our $J$, information about $\gamma$ in (\ref{5}) can be
obtained by comparing the two-point functions $\langle JJ \rangle$
and $\langle M \bar M \rangle$ or the three-point functions $\langle
\cO \bar \cO J \rangle$ and $\langle \cO \bar \cO N \rangle$. In
particular, the one-loop value of $\g$ can be found without doing
any divergent $x$-space Feynman integrals, as we explain in this
section.

\subsection{Anomalous dimension through superconformal two-point functions}\label{bos}

\subsubsection{The bosonic case}

In order to illustrate the idea of the method, we first consider the purely bosonic case where the ``current" $J$ satisfies the non-conservation condition (\ref{4'}). Dealing with arbitrary spin requires an efficient tool for traceless symmetrization of the vector indices or, equivalently, for  symmetrization of the spinor indices of $J$. Such a tool was introduced in \cite{DPPT}, where an light-like vector $z^\mu$, $z^2=0$ was used to project all the vector indices of a given tensor, thus automatically symmetrizing and suppressing the traces. Later on, in \cite{DP2} a version more suitable for the supersymmetric case (i.e., for handling spinor indices) was proposed. It consists in replacing $z^\mu$ by a pair of {\sl commuting} spinor variables $z^\a$, $\bz^{\da}$ as follows: $z^{\a\da} = z^\mu (\sigma_\mu)^{\a\da} = z^\a \bz^{\da}$. So, projecting all indices of $J$,
\begin{equation}\label{9'}
\hat{J} = z^{\alpha_{1}} \cdots z^{\alpha_{j}}\, J_{\alpha_{1}
\cdots \alpha_{j} \da_{1} \cdots \da_{j}} \, \bar{z}^{\da_{1}}
\cdots \bar{z}^{\da_{j}}
\end{equation}
makes the symmetrization manifest.

Let us now apply this tool for constructing the two-point function of $\hat J$. Conformal invariance fixes its form:\footnote{For comprehensive reviews of conformal symmetry see, e.g. \cite{FP}. Throughout this section the normalizations are not fixed, since we are only interested in ratios of two- and three-point functions.}
\begin{equation}\label{10'}
\langle \hat{J}(1) \hat{J}(2)  \rangle =C\,
\frac{(\hat{x}_{1{2}}
\hat{x}_{2{1}})^{j}}
{(x_{12}^{2}x_{21}^{2})^{(d+j)/2}} \,,
\end{equation}
where $x_{12}=x_1 - x_2$, $d=2+j +\g$ is the conformal dimension of the operator $J$ and
\begin{equation} \label{9''}
\hat{x}_{1{2}} = z^\a_{1} (x_{12})_{\alpha \da} \bar{z}^{
\da}_{2}\,, \qquad
\hat{x}_{2{1}} = z^{\beta}_{2}  (x_{21})_{\beta \dot{\beta}}
\bar{z}^{\dot{\beta}}_{1} \,.
\end{equation}
The denominator in (\ref{10'}) contains the square $x^2_{12}= (x^\mu_{12})^2=x^2_{21}$, but it is written in this way to facilitate the supersymmetrization below. Notice that we have used  two {\it independent} sets of commuting spinors,  ($z_{1}$, $\bar{z}_{1}$) for $J(1)$ and ($z_{2}$, $\bar{z}_{2}$) for $J(2)$ to project the spinor indices. For the reader more familiar with the vector index notation, we mention the case of a current of spin $j=1$ with canonical dimension $d_0=3$, for which
\begin{equation}\label{1003}
  \langle J^\mu(x_1) J^\nu(x_2) \rangle  = \frac{C}{2}\,
\frac{x_{12}^2\delta^{\mu\nu} - 2 x_{12}^\mu x_{12}^\nu}{(x_{12}^2)^{4 + \gamma}}\,.
\end{equation}
Switching to two-component spinor notation and projecting the indices, we obtain (\ref{10'}).

Next, we wish to compute the two-point function of the divergence $\pa^{\a\da}J_{\a\ldots\, \da\ldots}\ $. Defining
\begin{equation}\label{1005}
\check{\pa} = \frac{\partial}{\partial z^\a} \pa^{\a\da} \frac{\partial}{\partial \bar z^{\da}}\,,
\end{equation}
we find
\begin{equation}\label{12000}
  \check\pa \hat J = j^2\, z^{\a_2}\cdots z^{\a_j}\, \pa^{\a\da}  J_{\a\a_2\cdots\a_j\ \da\da_2\cdots\da_j} \bz^{\da_2}\cdots \bz^{\da_j}\,.
\end{equation}
Further, differentiating eq.\,(\ref{10'}) at both points, using
(\ref{spinor-vector1}), (\ref{spinor-vector2}), and setting
$z_1=z_2$, $\bar z_1=\bar z_2$, we easily obtain
\begin{equation}\label{1006}
  \check{\pa}_1 \check{\pa}_2 \langle  \hat{J}(1) \hat{J}(2)  \rangle =j^4 g^2 \langle  \hat{K}(1) \hat{K}(2)  \rangle  =4Cj^2  \g[\g(j^2+1)+(j+1)^2]\, \frac{(\hat{x}_{1{2}}
\hat{x}_{2{1}})^{j-1}}
{(x_{12}^{2}x_{21}^{2})^{(d+j)/2}} \,,
\end{equation}
where we have inserted the descendant $K$ (\ref{4'}).

Now, consider the ratio
\begin{equation}\label{1007}
 j^4 g^2\hat{x}_{1{2}}
\hat{x}_{2{1}}\ \frac{ \langle  \hat{K}(1) \hat{K}(2)  \rangle}{\langle \hat{J}(1) \hat{J}(2)  \rangle} =  4 j^2 \g[\g(j^2+1)+(j+1)^2] = 4 j^2(j+1)^2\g_1 g^2  + O(g^4)\,.
\end{equation}
In practical terms, we have gained an order in $g^2$. If we wished to compute, e.g., $\gamma_1$ starting from eq.\,(\ref{10'}), we would have to evaluate {\it one-loop} divergent integrals involved in the two-point function of $J$. However, using the ratio (\ref{1007}) instead, it is sufficient to compute the {\it tree-level} two-point functions of the current and of its divergence. It is precisely this trick that Anselmi  \cite{Anselmi} applied to the current in the Konishi multiplet in order to obtain its one-loop anomalous dimension, without any loop integrals.

\subsubsection{The supersymmetric case}

Supersymmetry provides another way of doing this calculation. Indeed, the space-time derivative in (\ref{1006}) amounts to two consecutive superspace differentiations (or supersymmetry transformations), see (\ref{2'}). Instead, we could differentiate just once as indicated in (\ref{4}), and still obtain a ``superdivergence" proportional to $g$. To this end we need the supersymmetrization of the bosonic two-point function (\ref{10'}) which can be obtained as follows \cite{Park:1999pd}. First, we introduce left- and right-handed chiral bases in superspace,
\begin{equation} \label{6} x_{L}^{\alpha \da} = x^{\alpha \da} + 2i \theta^{A\,\alpha}
\bar{\theta}_A^{\da} \,, \qquad x_{R}^{\alpha \da} = x^{\alpha \da} - 2i \theta^{A\,\alpha}
\bar{\theta}_A^{\da} \,.
\end{equation}
Then we form the supersymmetric invariant
\begin{equation}\label{7}
 x^{\alpha \da}_{1\bar{2}} = x_{1_L}^{\alpha \da} - x_{2_R}^{\alpha
\da} - 4i \theta_{1}^{A\,\alpha}
\bar{\theta}_{2\, A}^{\da}
\end{equation}
and replace $x_{12}$ and $x_{21}$ in (\ref{10'}) to obtain
\begin{equation}\label{10}
\langle \hat{J}(1) \hat{J}(2)  \rangle =C\,
\frac{(\hat{x}_{1\bar{2}}
\hat{x}_{2\bar{1}})^{j}}
{(x^{2}_{1\bar{2}} x^{2}_{2\bar{1}})^{(d+j)/2}} \,.
\end{equation}
The next step is the spinor differentiation of both sides according
to (\ref{4}). We are only interested in $D^\a_A J_{\a\ldots}$ and
$\bD^{A\, \da} J_{\da\ldots}$ at $\q=\bq=0$, so (\ref{1008}) is
reduced to just $\frac{\partial}{\partial \theta}$ $\frac{\partial}{\partial \bar\theta}$. Thus, calculating $\langle
D^\a_A J_{\a\ldots}(1)  \bD^{B\, \da} J_{\da\ldots}(2) \rangle$
amounts to extracting the $\q^{C}_1\bq_{2 C}$ term in the expansion
of (\ref{10}). It is clear that such terms only originate from
$x_{1\bar 2}$ but not from $x_{2\bar 1}$. Defining
\begin{equation}
\check{D}_A = D^\a_A \frac{\partial}{\partial z^\a} \,, \qquad \check{\bar{D}}^A = \bar{D}^{A\,\da}\frac{\partial}{\partial \bar{z}^{\da}}\,,
\end{equation}
we find
\begin{equation}\label{11}
\check{D}_{1\, A}\check{\bar{D}}_2^B \langle \hat{J}(1) \hat{J}(2)
 \rangle|_{\q=\bq=0}  =  -2iC \delta_{B}^{A} j (j+1)\gamma\ \frac{
(\hat x_{12})^{j-1} (\hat x_{21})^{j}}{(x^{2}_{1{2}})^{d+j}}\,.
\end{equation}
Recalling that the spinor divergence of $J$ defines its descendant $M$ (\ref{4}), we find the ratio of  two-point functions
\begin{equation}\label{12}
 \left.  g^2\ \frac{\hat x_{12} \langle \hat{M}^{A}(1) \hat{\bar M}_B(2) \rangle}{\langle \hat{J}(1) \hat{J}(2)  \rangle}\right|_{\q=\bq=0}
 = -2i \delta_{B}^{A} j (j+1) \gamma \,.
\end{equation}
We see that both sides of this equation are proportional to $g^2$, as expected.  The conclusion is that at the lowest order in perturbation theory, $\gamma(g^2) = \gamma_1 g^2 + \ldots\ $,  we can avoid computing loop corrections to the two-point functions. Instead, we can calculate them at tree level and then evaluate the ratio (\ref{12}):
\begin{equation}\label{1200}
 \left. \frac{\hat x_{12}\,\langle \hat{M}^{A}(1) \hat{\bar M}_B(2) \rangle_{g^0}}{\langle \hat{J}(1) \hat{J}(2)  \rangle_{g^0}}\right|_{\q=\bq=0}
 = -2i \delta_{B}^{A} j (j+1)\ \gamma_1 \,.
\end{equation}

\subsection{An alternative method based on three-point functions}

The two-point function method for obtaining the one-loop anomalous dimension of $J$ through tree-level calculations described above has one technical but important drawback. The structure of the classical descendants $K$ in (\ref{1007}) and $M$ in (\ref{1200}) is rather involved (see Section \ref{ICD}), leading to
complicated quadruple sums (see (\ref{21})). It turns out that a much more efficient way is to compare the three-point functions $\langle \cO \bar \cO J \rangle$ and $\langle \cO \bar \cO N \rangle$, where $\cO$ is the half-BPS operator (\ref{505'}) and $N$ is the second-level descendant of $J$ (recall (\ref{4001}) and (\ref{4002})). This ratio is proportional to $\gamma =g^2\gamma_1 + \ldots\ $, therefore it can be evaluated at tree level, thus giving $\gamma_1$. The advantage of this method is the linear insertion of the descendant $N$, as opposed to the quadratic expression in $M$ in the two-point function approach. This allows us to avoid the complicated quadruple sums in (\ref{21}).

\subsubsection{The bosonic case}\label{boscas}

Let us illustrate the method by a simple bosonic example. Let $\cO = \phi^2$ be a bilinear composite operator made out of a free complex scalar field $\phi(x)$. A current satisfying (\ref{3}) for $j=1$ \footnote{We remind the reader that in the $\cN=4$ SYM theory the semishort operators can have only even spin. Nevertheless, the simple case of spin $j=1$ can serve as a model for constructing conformal, and later on superconformal three-point functions. } can be realized as another bilinear,
\begin{equation}\label{5'}
  J^\mu(x) = \frac{i}{2}(\bar\phi \pa^\mu \phi - \phi \pa^\mu \bar\phi)\,.
\end{equation}
Indeed, $\pa_\mu J^\mu = 0$ as a corollary of the free field equation $\square\phi=0$. The three-point function $\langle \cO(1)\, \bar\cO(2)\, J^\mu(3)\rangle $ can be obtained in the following way. First, using the free scalar propagator\footnote{The standard normalization will be restored in Section \ref{CCLL}.}
\begin{equation}\label{5'''}
  \langle\phi(1)\,\bar\phi(2)\rangle = \frac{1}{x^2_{12}}\,,
\end{equation}
we form the scalar three-point function
\begin{equation}\label{5''}
  \langle \phi^2(1)\, \bar\phi^2(2)\, \bar\phi\phi(3)\rangle = \frac{1}{x^2_{12}x^2_{13}x^2_{23}}\,.
\end{equation}
Then we differentiate it at point 3 according to the structure of the current (\ref{5'}), thus obtaining
\begin{equation}\label{6001}
  \langle \cO(1)\, \bar\cO(2)\, \hat J(3)\rangle = \frac{i}{x^4_{12}}\, Y^2\,\hat Y \,.
\end{equation}
Here
\begin{equation}\label{7001}
  \hat  Y = \frac{\hat x_{31}}{x^2_{31}} - \frac{\hat x_{32}}{x^2_{32}}\,, \qquad Y^2 = \frac{x^2_{12}}{x^2_{13}x^2_{23}}\,,
\end{equation}
is a three-point vector which is invariant at points 1 and 2 and transforms covariantly at point 3 under conformal boosts. Note that we have projected the Lorentz indices with the auxiliary variables $z,\bar z$.

The generalization to a $J$ with arbitrary spin $j$ and conformal dimension $d$ is straightforward:
\begin{equation}\label{8}
  \langle \cO(1)\, \bar\cO(2)\, \hat J(3)\rangle = \frac{C}{x^4_{12}}\, (Y^2)^{\frac{d-j}{2}}\,\hat Y^j \,.
\end{equation}
In the particular case $d=j+2$ (canonical dimension) the three-point function (\ref{8}) automatically satisfies the conservation law (\ref{3}) at point 3 (actually, this provides a way to construct (\ref{8}) for $d=j+2$, by applying to point 3 a differential operator giving a conserved tensor, see below). Notice that in (\ref{8}) we have assumed that the operator $\cO$ keeps its canonical dimension two, otherwise the prefactor $1/x^4_{12}$ would have to be modified.

Conformal covariance allows us to choose a special frame in which (\ref{8}) is greatly simplified. Using conformal boosts we can set $x_2=\infty$, after which we find
\begin{equation}\label{9}
  \lim_{x_2\to\infty}\ x^4_2\,\langle \cO(1)\, \bar\cO(2)\, \hat J(3)\rangle =  C\,(x^2_{31})^{-\frac{d+j}{2}}\, \hat x_{31}^{j}\,.
\end{equation}
Further, differentiating eq.\,(\ref{9}) with $\check\pa_3$, we easily obtain
\begin{equation}\label{13'}
  \lim_{x_2\to\infty}\ x^4_2\,\langle \cO(1)\, \bar\cO(2)\, \check\pa J(3)\rangle = C\,j(j+1)(j+2-d)(x_{31}^2)^{-\frac{d+j}{2}}\; (\hat x_{31})^{j-1}\,.
\end{equation}
This confirms the conservation when $d=j+2$, as stated above. Note that the limit and the derivative in (\ref{13'}) commute, because if the derivative hits a negative power of $x_{32}$ in (\ref{8}), it makes it vanish even faster in the limit.

Finally, let us consider the ratio of the three-point functions (\ref{13'}) and (\ref{9})
\begin{equation}\label{14}
 \lim_{x_2\to\infty}\  \frac{{\hat x_{31}}\langle \cO(1)\, \bar\cO(2)\, \check\pa J(3)\rangle}{\langle \cO(1)\, \bar\cO(2)\,  J(3)\rangle} = -{j(j+1)}\gamma \,.
\end{equation}
As in the two-point case, it is proportional to the anomalous dimension $\gamma(g^2) = \gamma_1 g^2 + O(g^4)$. This implies that the left-hand side of eq.\,(\ref{14}) must be proportional to $g^2$. However, according to eq.\,(\ref{4'}) the space-time derivative, i.e. the anticommutator of two spinor derivatives at point 3 produces a single factor of $g$. The missing factor should then be supplied by a one-loop correction to the correlator in the numerator of  (\ref{14}), so there would be no gain compared to a traditional calculation of $\gamma$. We can do better if instead we use the commutator of two spinor derivatives at point 3, according to the definition of the descendant $N$ (\ref{4001}). To this end we need to find the supersymmetric generalization of the three-point functions above.

\newpage

\subsubsection{The supersymmetric case}

When constructing the supersymmetric analog of the bosonic three-point function (\ref{8}), we can keep points 1 and 2 at $\q=\bq=0$. However, we wish to differentiate $\langle\cO\bar\cO J\rangle$ twice with spinor derivatives at point 3, so we need to reconstruct the $\q$ dependence at this point. There exist general methods for this \cite{Park:1999pd}, but here we can use a simpler construction.

First of all, we note that a superconformal covariant with two half-BPS points and one generic point is uniquely determined by its lowest (bosonic) component. We can find this covariant by repeating the bosonic procedure indicated at the beginning of subsection \ref{boscas}.
The analog of the scalar three-point function (\ref{5''}) is obtained by putting $\cO=\T(W^{12}W^{12})$ at point 1 and  $\bar\cO=\T(W_{12}W_{12})$ at point 2. At point 3 we need an $SU(4)$ singlet of dimension two, the so-called Konishi operator $\T(W^{AB}W_{AB})$. So, our starting point is the three-point function
\begin{equation}\label{5007}
  \langle\T(W^{12}W^{12})(1)\ \T(W_{12}W_{12})(2)\ \T(W^{AB}W_{AB})(3)\rangle
\end{equation}
at tree level. To construct it we need to connect the three points with free field propagators as shown in the figure:
\\
\\
\begin{minipage}{\textwidth}
\begin{center}
\includegraphics[width=0.28\textwidth]{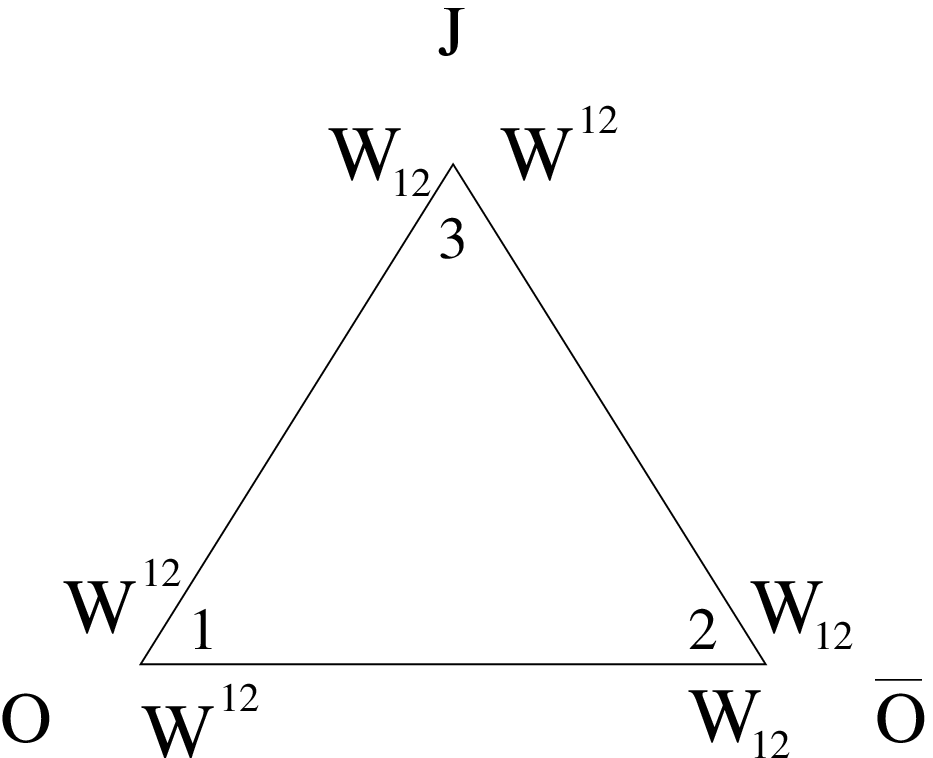}
\end{center}
\end{minipage}
\\
\\
\\
\\
 The Grassmann analytic superfields $W^{12}$ at point 1 can only see their conjugates $W_{12}=W^{34}$ via the propagator
\begin{equation}\label{507}
 \langle W^{12}(X, \q^{1,2},\bq_{3,4})\ W_{12}(\bar Z, \zeta^{3,4},\bar\zeta_{1,2})  \rangle = (X - \bar Z - 4i(\theta^{1}\bar\zeta_1 + \theta^{2}\bar\zeta_2) + 4i(\zeta^{3}\bar\theta_3 + \zeta^{4}\bar\theta_4))^{-2}\,.
\end{equation}
Here $X$ is the analytic basis coordinate (\ref{503}) and $\bar Z$
is its antianalytic counterpart obtained by conjugation. The
$\q,\zeta$ terms in (\ref{507}) make the whole expression invariant
under supersymmetry, in close analogy with the chiral case
(\ref{7}). In fact, we do not even need them since we have set the
odd coordinates at points 1 and 2 to zero. Further, we have to
create the structure of the operator $J$ at point 3 by
differentiating the propagators $1\to3$ and $2\to3$, just as we did
to obtain (\ref{6001}) from (\ref{5''}). To this end we need the
supersymmetric analog of (\ref{5'}) for arbitrary spin $j$, which is
worked out in Section \ref{TTO}, eq.\,(\ref{13}) and thereafter. The
superspace version of (\ref{13}) is obtained by replacing
$\phi^{AB}$ by $W^{AB}$, $\lambda^A_\a$ by $-i\sqrt{2}/6\ D_{B\, \a}
W^{AB}$ and $F_{\a\b}$ by $-i\sqrt{2}/24\ D_{A\, \a}D_{B\, \b}
W^{AB}$. The result resembles (\ref{8}), with the three-point
covariant $Y$ (\ref{7001}) replaced by its superconformal completion
${\cal Y}$,
\begin{equation}\label{508}
  \langle \cO(1)\, \bar\cO(2)\, \hat J(3)\rangle = \frac{C}{x^4_{12}}\, ({\cal Y}^2)^{\frac{d-j}{2}}\,\hat {\cal Y}^j \,
\end{equation}
(we recall that the odd coordinates at points 1 and 2 have been set
to zero, so there is no difference between $x_{1}$ and $X_1$ or
$x_2$ and $\bar X_2$).

This expression is drastically simplified in the limit $\lim_{x_2
\to \infty}\, x_2^4\, \langle \cO(1)\, \bar\cO(2)\, \hat
J(3)\rangle$. Indeed, the contributions from the second and the
third terms in the right-hand side of (\ref{13}) will vanish in this
limit. For example, the term $\langle W_{12}(2)\, \lambda^A(3)
\rangle$ is given by
\begin{equation}\label{509}
\langle W_{12}(2) \, D_B^\a W^{AB}(3) \rangle =-2i \ [\delta^A_2 (\sigma^\mu \bar \theta^1)^\alpha - \delta^A_1 (\sigma^\mu \bar \theta^2)^\alpha ] \frac{\partial}{\partial X^\mu_{3}}\ \frac{1}{(x_2-X_3)^2}\,,
\end{equation}
where $\bq$ is at point 3 and we have used (\ref{1008}) rewritten in the analytic basis (\ref{503}).  This term has to be multiplied by $x_2^2$ before taking the limit $x_2 \to
\infty$ (the other factor $x_2^2$ neutralizes the propagator
$\langle W^{12}(1)W_{12}(2)\rangle$). Even so, it behaves like
$x_2^{-1}$ and thus vanishes. The contribution from the $F$ terms in
(\ref{13}) vanishes for similar reasons. Moreover, in this limit the
covariant $Y$ (\ref{7001}) is reduced to just $x_{31}/x^2_{31}$,
whose supersymmetrization is obtained by replacing $x_{3}$ by $\bar
X_3$, according to the analyticity property of the propagator
$\langle W^{12}(1) W_{12}(3)\rangle$ at point 3. So, finally,
\begin{equation}\label{510}
 \lim_{x_2 \to \infty}\, x_2^4\, \langle \cO(x_1=0)\, \bar\cO(2)\, \hat J(3)\rangle = C (\bar X_3^2)^{-\frac{d+j}{2}}\, (\hat{\bar{X}}_3)^j\,.
\end{equation}
Note that throughout the process of constructing the three-point function we implicitly kept $d$ at its canonical value $j+2$, because we were using free-field propagators. Nevertheless, the result (\ref{510}) is valid also in the case of anomalous dimension.

Our last step is to differentiate the correlator (\ref{510}) at point 3 according to (\ref{4002}), in order to produce the second-level descendant $N^A_B$. We can pick, for instance, the $SU(4)$ projection $N^1_1$ given by the differential operator
\begin{equation}\label{511}
([\check{\bar D}^1,\check D_1] - 1/4\, [\check{\bar D}^C,\check D_C])\hat J|_{\theta=0} = 2g^2 \hat N^1_1|_{\theta=0}\,.
\end{equation}
When applied to (\ref{510}), these derivatives simplify even
further.\footnote{Taking the limit first and differentiating
afterwards is legitimate, because if any of the derivatives in
(\ref{511}) hits  $x_{23}$, it creates additional negative powers of
$x_2$.} Indeed, in the antianalytic basis (the conjugate of
(\ref{503})) the derivatives $D_{1,2}$ and $\bar D^{3,4}$ are short,
i.e., they annihilate $\bar X_3$; using (\ref{2'}), we can reduce
(\ref{511}) to just the space-time derivative $i\check\pa_{\bar
X_3}$. So, the calculation is trivial:
\begin{equation}\label{512}
 g^2\, \lim_{x_2 \to \infty}\, x_2^4\, \langle \cO(x_1=0)\, \bar\cO(2)\, \hat N^1_1(3)\rangle = -iCj(j+1)\g (\bar X_3^2)^{-\frac{d+j}{2}}\, (\hat{\bar{X}}_3)^{j-1}\,.
\end{equation}

Finally, the ratio of the three-point functions (\ref{512}) and (\ref{510}) determines the anomalous dimension,
\begin{equation}\label{513}
  g^2\hat{x}_3\,  \lim_{x_2 \to \infty} \frac{\langle \cO \bar\cO \hat N^1_1 \rangle|_{\theta=0}}{\langle \cO \bar\cO \hat J \rangle|_{\theta=0}} = -i\, j(j+1) \gamma\,.
\end{equation}

\newpage

\section{Determining $J$ and $M$ at the classical level}\label{CCL}

\subsection{Twist-two operators in the free theory}\label{TTO}

In this subsection we construct the twist-two operators $\hat{J}$ out of the elementary component fields of the $\cN=4$ SYM theory (see (\ref{504})). This result has already been obtained in \cite{BHR} using $\cN=4$ analytic superspace. Here we give an alternative derivation in terms of component fields instead of superfields. Our experience has shown that the component approach is more efficient, especially for constructing the descendants $M$ and $N$ of $J$. We make use of the formalism of bilocal operators proposed in \cite{DPPT} (see also its supersymmetric version in \cite{DP2}). In the free theory ($g=0$) we write down a composite gauge invariant operator where the space-time points of the constituent fields have been split apart (the limit  $x_a\to x_b \equiv x $  will be taken at the end):
\begin{eqnarray}\label{13}
\hat{J} &=& P_{j}(\hat{\partial}_{a},\hat{\partial}_{b})
\T(\phi^{AB}(x_a) \phi_{AB}(x_b)) \nonumber \\ &+&
R_{j-1}(\hat{\partial}_{a},\hat{\partial}_{b})  \T(
 \hat{\lambda}^{A}(x_a) \hat{\bar{\lambda}}_{A}(x_b)) \nonumber \\ &+&
S_{j-2}(\hat{\partial}_{a},\hat{\partial}_{b}) \T(\hat{F}(x_a)
\hat{\bar{F}}(x_b) )\,.
\end{eqnarray}
Here $\hat{\partial} = z^\a \partial_{\alpha\da}
\bar{z}^{\da}$, $\hat{\lambda}=z^\a\lambda_\a$,
$\hat{F}=z^\a z^{\beta}F_{\alpha\beta}$, etc., and
$P_{j}(\hat{\partial}_{a},\hat{\partial}_{b})$, etc. denote bilocal
differential operators in the form of homogeneous polynomials,
e.g.
\begin{equation}
P_{j}(\hat{\partial}_{a},\hat{\partial}_{b}) = \sum_{k=0}^{j} p_k
(\hat{\partial}_{a})^k (\hat{\partial}_{b})^{j-k} \label{14pk}
\end{equation}
and similarly for $R$ and $S$ (note that we use the same auxiliary variables $z,\bar z$ for projecting $\pa_a$ and $\pa_b$). The conservation conditions (\ref{2}), translated into component language, take the following bilocal form:
\begin{equation}\label{15bilocal}
(\check{Q}_a + \check{Q}_{b}) \hat{J} = (\check{\bar{Q}}_a + \check{\bar{Q}}_{b})\hat{J} = 0\,.
\end{equation}
Here we are using the free ($g=0$) on-shell supersymmetry transformations (\ref{susy}) generated by the supercharges $Q,\bar Q$. Applying them to $J$ and using the free version of the field equations (\ref{eqm}) and their corollaries, it is not hard to find four differential equations for the polynomials $P$, $R$ and $S$
(for brevity we denote $a=\hat{\partial}_a$, $b=\hat{\partial}_b$).
\begin{eqnarray}\label{16}
 4i \frac{\partial P}{\partial
b} - \left( a \frac{\partial R}{\partial a} +R \right) &=& 0 \nonumber \\
 4i \frac{\partial P}{\partial
a} + \left( b \frac{\partial R}{\partial b} +R \right) &=& 0\nonumber \\
\frac{\partial R}{\partial b} -2i \left(a \frac{\partial S}{\pa a}
+2S \right) &=&
0 \nonumber \\
\frac{\partial R}{\partial a} +2i \left(b \frac{\partial S}{\pa b}
+2S \right) &=& 0\,.
\end{eqnarray}
Their solutions can be obtained\footnote{In \cite{DPPT} the
homogeneous polynomial $P$ is rewritten in terms of a single
variable $(\partial_a-\partial_b)/(\partial_a + \partial_b)$ and the
corresponding differential equation is identified with that for a
Gegenbauer polynomial. This is the standard form for twist-two
operators in the QCD literature. For our purposes, however, it is
preferable to have the explicit distribution of derivatives on the
two fields, as in (\ref{14pk}). Such a form, only for scalar constituents
appears in \cite{Sezgin}. } by substituting the expansions (\ref{14pk}), etc.
and solving the resulting recurrence relations. We find, in agreement with \cite{BHR},
\begin{eqnarray}\label{17}
  p_k &=& (-1)^k
  \begin{pmatrix}
    j \\
    k
  \end{pmatrix}^2
  \nonumber\\
  r_k &=& 4i (-1)^{k}
  \begin{pmatrix}
    j \\
    k+1
  \end{pmatrix}  \begin{pmatrix}
    j \\
    k
  \end{pmatrix}
  \nonumber\\
  s_k &=& 2 (-1)^k
  \begin{pmatrix}
    j \\
    k+2
  \end{pmatrix}  \begin{pmatrix}
    j \\
    k
  \end{pmatrix}\,.
\end{eqnarray}
It can easily be verified that $J$ is real.

We remark that the point permutation symmetry in the scalar sector in (\ref{13}) requires that $p_{j-k}=p_k$, which is only compatible with (\ref{17}) if $j$ is even. This is a property of the maximally supersymmetric $\cN=4$ SYM theory. In theories with less supersymmetry twist-two operators with odd spin do exist.

\subsection{Interaction and classical descendants}\label{ICD}

The interacting field theory version of the operators $J$ is
obtained by replacing the projected derivatives $\hp$ by covariant
ones, $\hat {\cal D}$. When checking the conservation conditions (\ref{2}) we made
use of the free field equations. This allowed us to drop a great
number of terms, which become active in the interacting theory, thus
giving rise to descendants like in (\ref{4}). Working out the full
expression with the help of the non-linear supersymmetry transformations (\ref{susy}) and field equations (\ref{eqm}), we find
\begin{eqnarray}\label{18}
\hat{M}_{C} = \check{Q}_{C}\hat{J} = \T \Big[ && \sum_{k=2}^{j}
\sum_{m=2}^{k} B_{j}(k,m) \hat{\cal D}^{j-k} \hat{\lambda}^{A}
\lbrack \hat{\cal D}^{k-m} \phi_{AC}, \hat{\cal D}^{m-2}
\hat{\bar{F}} \rbrack  \nonumber \\ && + \sum_{k=1}^{j}
\sum_{m=0}^{k-1} C_{j}(k,m) \hat{\cal D}^{j-k}
\hat{\bar{\lambda}}_{A} \lbrack {\hat{{\cal D}}}^{k-1-m} \phi_{BC},
\hat{{\cal D}}^{m} \phi^{AB}\rbrack \nonumber \\&& + \sum_{k=3}^{j}
\sum_{m=2}^{k-1} D_{j}(k,m) \, \hat{{\cal D}}^{k-1-m}
\hat{\bar{\lambda}}_{C} \lbrack \hat{{\cal D}}^{j-k} F, \hat{{\cal
D}}^{m-2} \bar{F} \rbrack \nonumber \\ && + \sum_{k=2}^{j}
\sum_{m=0}^{k-2} E_{j}(k,m) \hat{\cal D}^{j-k}
\hat{\bar{\lambda}}_{C} \{ \hat{\cal D}^{m} \hat{\bar{\lambda}}_{A},
\hat{\cal D}^{k-2-m} \hat{\lambda}^{A} \}\ \Big] \,.
\end{eqnarray}
Here
\begin{eqnarray}
&& B_{j}(k,m) = 4\sqrt{2}(m-1)
{\scriptsize \begin{pmatrix} k \\ m \end{pmatrix}}
{\scriptsize \begin{pmatrix} j \\ k \end{pmatrix}}  \nonumber \\
&&\times\bigg\{ (-1)^{k+m} {\scriptsize\begin{pmatrix} j+1 \\ k-m
\end{pmatrix}}
  -(-1)^{k}
{\scriptsize \begin{pmatrix} j+1 \\ k \end{pmatrix}}
 - (-1)^{m}
{\scriptsize \begin{pmatrix} j+1 \\ m-1 \end{pmatrix}}
\bigg\} \ \ \  k=2...j, \,  m=2...k  \label{19-1}\\
&& C_{j}(k,m) = \left \{ \begin{array}{ll} \sqrt{2} \frac{(m+1)(m+2)}{(j-k+1)(j-k)}
B_{j}(k+1,m+2)     \\
8(m+1)^2 {\scriptsize \begin{pmatrix} j \\ m+1 \end{pmatrix}} \bigg \{ 1-(-1)^m{\scriptsize \begin{pmatrix} j+2 \\ m+2 \end{pmatrix}} \bigg \}  \end{array} \right .  \begin{array}{lll} k=1...j-1 , \, m=0...k-1 \\ \\ k=j , \, m=0...k-1 \end{array} \label{19-2} \\
&& D_{j}(k,m) = -\frac{1}{\sqrt{2}} \frac{j-k+1}{k-m} B_{j}(k-1,m) \quad  \quad \quad \quad \quad  \quad \ \ \ \, k=3...j , \, m=2...k-1 \label{19-3} \\
&& E_{j}(k,m) = i \sqrt{2} \frac{m+2}{j-k+1} B_{j}(j-k+m+2,m+2) \quad \quad \ k=2...j, \, m=0...k-2
 \label{19-4} \,
\end{eqnarray}
and vanish for all other values of $k$ and $m$. Note that $C_j(j,m)=\lim_{k \to j} C_j(k,m)$.

As stated in \cite{BHR}, $M$ is subject to the constraint
\begin{equation}\label{19'}
\check{\bar{Q}}^{D}\hat{M}_{C}-\frac{1}{4}\delta^{D}_{C}\check{\bar{Q}}^{A}\hat{M}_{A}
= O(g) \,,
\end{equation}
that we have checked explicitly. We have also found that the general solution of eq.\,(\ref{19'}) is given by
eqs.\,(\ref{19-2})--(\ref{19-4}) (except for $C_j(j,m)$, which remains undetermined), with $B_{j}$ satisfying the following symmetry property:
\begin{equation}\label{19-5}
(k-m)B_{j}(k,m) = (j-k+1)B_{j}(j-k+m+1,m)\,.
\end{equation}

\section{Calculation of the anomalous dimension of the twist-two operators}\label{CCLL}

\subsection{The two-point way}

In this subsection we work out the two point functions
$\langle J\, J\rangle$ and $\langle M \, \bar{M} \rangle$ at tree level, in order to compute the one-loop anomalous dimension $\g_1$ according to (\ref{1200}).\\

The calculation using the expression (\ref{13}) for $J$ and the
propagators (\ref{scalarprop})--(\ref{gluonprop}) is straightforward. We find (here, e.g., $x_1\equiv x$, $x_2=0$)
\begin{equation}\label{20}
\langle J\, J\rangle =  \frac{2^{2j+2} N_c}{(4 \pi^2)^2} \frac{\hat{x}^{2j}}{(x^{2})^{(2j+2)}}
 (12S_{p}+S_{r}+S_{s}) \,,
\end{equation}
where (see (\ref{17}) for the expressions of $p_{k}$, $r_{k}$, and
$s_{k}$)
\begin{eqnarray} \label{20'} S_{p} &=& \sum_{k=0}^{j}
\sum_{m=0}^{j} p_k p_m (j-m+k)! (j-k+m)! \nonumber \\
S_{r} &=& \sum_{k=0}^{j-1}
\sum_{m=0}^{j-1} r_k r_m (j-m+k)! (j-k+m)! \nonumber \\
S_{s} &=& \sum_{k=0}^{j-2} \sum_{m=0}^{j-2} s_k s_m (j-m+k)!
(j-k+m)!\,
\end{eqnarray}

The calculation of $\langle M \, \bar{M} \rangle$ involves quadruple
sums and hence is rather complicated:
\begin{eqnarray}
  \langle M^A\, \bar M_B\rangle &=&  -i \delta^A_B\frac{2^{2j} N_c^2}{(4 \pi^2)^3} \frac{\hat{x}^{2j-1}}{(x^{2})^{(2j+2)}} \sum_{k=0}^j \sum_{k'=0}^j \sum_{i=0}^k \sum_{i'=0}^k \Big\{ \nonumber\\
  &&  6 B(k,i) B(k',i') (i+i'-2)!(k+k'-i-i')!(2j-k-k'+1)! \nonumber\\
  &&  + C(k,i) C(k',i') [9(i+i')!(k+k'-i-i'-2)!  \nonumber\\
  && +6(k'-i'+i-1)!(k-i+i'-1)!] (2j-k-k'+1)!  \nonumber\\
  && +  4D(k,i) D(k',i') (i+i'-2)!(k+k'-i-i'-1)!(2j-k-k'+2)!       \nonumber\\
  &&  - E(k,i) E(k',i') [4(i+i'+1)!(2j-k-k'+1)!  \nonumber\\
  && +(j-k'+i+1)!(j-k+i'+1)!] (k+k'-i-i'-3)!\Big\}\,    \label{21}
\end{eqnarray}
(here we assume that all factorials vanish for negative values of their arguments).

We have verified numerically that the ratio (\ref{1200}) does indeed reproduce the known value of $\gamma_1$ (see (\ref{33})). However, the three-point ratio (\ref{513}) provides a much simpler way to obtain the same result.

\subsection{The three-point way}

Here we evaluate the tree-level three-point correlators in
(\ref{513}) by using the explicit realizations of the operators
$\cO$, $J$ and $N$ in terms of component fields. We already have $J$
in that form ((\ref{13}), (\ref{17})), but working out the complete
expression for $N$ is a long and tedious calculation. We have to
start with the expression for $M$ (\ref{18})-(\ref{19-4}), apply to
it the supercharges from (\ref{4001}) and use the equations of
motion. Fortunately, this task is drastically simplified at tree
level. The operator $\cO=\T(\phi^{12}\phi^{12})$ involves only scalars, which can only talk to the
conjugate scalar in $J$ or $N$ via a free
propagator. This allows us to keep in $N$ only terms quadrilinear in
the scalar fields,

\begin{eqnarray}\label{N}
N_A^B &=& i \sum_{k=1}^{j} \sum_{m=0}^{k-1} \sum_{l=0}^{m} {\rm Tr} \bigg
\{ \bigg (
 (j+1+m-2l)C(j-m,j-k)+(2l-m)\sqrt{2}B(k+1,m+2) \bigg )\nonumber \\ && \times \ {\scriptsize \begin{pmatrix} m \\ l \end{pmatrix}} \lbrack \partial^{j-k} \phi^{CD},
\partial^{k-1-m} \phi_{AD} \rbrack \lbrack \partial^{l}
\phi^{B E}, \partial^{m-l} \phi_{C E} \rbrack \bigg \} -\frac{1}{4} \delta_A^B \{ {\rm trace} \},
\end{eqnarray}
and $B$ and $C$ were defined in (\ref{19-1}) and (\ref{19-2}) (recall that they vanish for ``illegal" values of their arguments). Next, we restrict (\ref{N}) to the projection $N^1_1$ as in (\ref{513}), and further to the terms containing only $\phi_{12}$ and $\phi^{12}$ (which can talk to the scalar fields in $\cO$ and $\bar\cO$). Finally, another major simplification is achieved in the limit
$x_2\to\infty$, where we can drop all terms with derivatives on
$\phi^{12}$, thus obtaining
\begin{eqnarray}
\lim_{x_2\to\infty} \ x_{2}^{4}\langle \cO\bar\cO J \rangle &=& -\frac{16N_c2^j\,j!}{(4\pi^2)^3} \ \frac{{\hat x}^j_{3}}{(x_3^2)^{j+1}} +O(g)\label{51} \\
\lim_{x_2\to\infty} \ x_{2}^{4}\langle \cO\bar\cO N_1^1 \rangle &=& i\frac{64 N_c^2 2^j\,j!}{(4\pi^2)^4}\  j(j+1)\ \left( \sum_{k=1}^{j+2} \frac{1}{k}  \right) \ \frac{{\hat x}^{j-1}_{3}}{(x_3^2)^{j+1}} + O(g)\,. \label{52}
\end{eqnarray}

Comparing the ratio of these dynamically determined three-point correlators to the kinematical prediction
(\ref{513}), we obtain our final result
\begin{equation}\label{33}
\gamma_1 =\frac{1}{-i\,j(j+1)}\ \frac{\hat x_3 \langle \cO\bar\cO N_1^1 \rangle_{g^0}}{ \langle \cO\bar\cO J \rangle_{g^0}} =  \frac{N_c}{\pi^2} \sum_{k=1}^{j+2} \frac{1}{k}\ .
\end{equation}
Thus, we have recovered the well-known
\cite{Lipatov:1loop,Dolan:1loop} value of the one-loop anomalous
dimension of twist-two operators of (even) spin $j$ by doing only tree-level
calculations.\footnote{In some references the value of $\g_1$
(\ref{33}) is divided by 2, which is due to a different convention
for the SYM coupling $g$.}

\section*{Acknowledgements}

It is a pleasure to thank L. Lipatov and V. Velizhanin for numerous and very stimulating discussions of higher spins in quantum field theory, which triggered our interest in this subject. We also acknowledge helpful discussions with M. Bianchi, P. Heslop and F. Riccioni.

\newpage

\section{Appendix: Conventions and definitions}

Our conventions are those of \cite{BDKM}. The $\cN=4$ SYM Lagrangian
in Minkowski space has the form
\begin{eqnarray}
{\cal L}_{{\cal N} = 4} = {\rm \T} \, \bigg\{ \!\!\!&-&\!\!\! \ft12
F_{\mu\nu} F^{\mu\nu} + \ft12 \left( {\cal D}_\mu \phi^{AB} \right)
\left( {\cal D}^\mu \phi_{AB} \right) + \ft{1}8 g^2 [\phi^{AB},
\phi^{CD}] [\phi_{AB}, \phi_{CD}]
\nonumber\\
&+&\!\!\! 2 i \bar\lambda_{\dot\alpha A} \sigma^{\dot\alpha
\beta}_\mu {\cal D}^\mu \lambda^A_\beta - \sqrt{2} g \lambda^{\alpha
A} [\phi_{AB}, \lambda_\alpha^B] + \sqrt{2} g
\bar\lambda_{\dot\alpha A} [\phi^{AB}, \bar\lambda^{\dot\alpha}_B]
\bigg\} \, . \label{lagrangian}
\end{eqnarray}
All fields are in the adjoint representation of the gauge group
$SU(N_c)$, and the generators and the structure constants satisfy the
relations
\begin{equation}
\T(t^{a} t^{b}) = \frac{1}{2}\delta^{ab}, \qquad f^{abd}f^{abe} = N_c\delta^{de}\,.
\end{equation}
The scalar fields $\phi^{AB}$ satisfy the reality condition
\begin{equation}
\phi_{AB} =  \overline{\phi^{AB}}  = \ft12 \ep_{ABCD}
\phi^{CD},
\end{equation}
where $\ep_{1234} = \ep^{1234} = 1$.\\

Spinor indices can be raised and lowered according to the rules
\begin{equation}
\lambda^\alpha = \ep^{\alpha \beta} \lambda_\beta \, ,
\qquad \bar\lambda_{\dot\alpha} = \ep_{\dot\alpha \dot\beta}
\bar\lambda^{\dot\beta} \, , \qquad \lambda_\alpha = \lambda^\beta
\ep_{\beta\alpha} \, , \qquad \bar\lambda^{\dot\alpha} =\bar\lambda_{\dot\beta} \ep^{\dot\beta \dot\alpha} \, ,
\end{equation}
where $\ep^{\alpha\beta} = - \ep_{\dot\alpha \dot\beta}$ and
$\ep^{12} = \ep_{12} = - \ep_{\dot 1 \dot 2} = - \ep^{\dot 1 \dot 2}
= 1$. We switch between spinor and vector notation using
\begin{equation}\label{spinor-vector1}
x^{\mu}=\frac{1}{2} \sigma^{\mu}_{\alpha \dot{\alpha}} x^{\alpha
\dot{\alpha}} \,, \qquad x_{\alpha \dot{\alpha}} =
\sigma^{\mu}_{\alpha \dot{\alpha}} x_{\mu} \,, \qquad x^2 = x_{\mu}
x^{\mu} = \frac{1}{2} x_{\alpha \dot{\alpha}} x^{\alpha
\dot{\alpha}}\,.
\end{equation}
Consequently, we have
\begin{equation}\label{spinor-vector2}
\partial^{\alpha\dot{\alpha}}x_{\beta\dot{\beta}} = 2
\delta^{\alpha}_{\beta} \delta^{\dot{\alpha}}_{\dot{\beta}}
\,,\qquad x_{\alpha \dot{\alpha}}x^{\alpha \dot{\beta}} = x^2
\delta_{\dot{\alpha}}^{\dot{\beta}} \,.
\end{equation}
The YM field strength is defined by the commutator of two covariant
derivatives which reads, in two-component spinor notation,
\begin{equation}\label{comcor}
  \lbrack {\cal D}_{\da\alpha}, {\cal D}_{\dot{\beta}\beta}
\rbrack = -ig (\ep_{\da\dot{\beta}}
F_{\a{\beta}} + \ep_{\a{\beta}}
\bar{F}_{\da\dot{\beta}} )\,,
\end{equation}
where
\begin{equation}
F^{\mu \nu} (\sigma_{\mu})^{\da\alpha}
(\sigma_{\nu})^{\dot{\beta}\beta} = F^{\alpha \beta \da
\dot{\beta}} = \ep^{\alpha \beta} \bar{F}^{\da
\dot{\beta}} + \ep^{\da \dot{\beta}} F^{\alpha
\beta}\,,
\end{equation}
or inversely,
\begin{equation}
\bar{F}^{\da}_{\dot{\beta}} = \frac{i}{2} F^{\mu \nu}
(\bar{\sigma}_{\mu \nu})^{\da}_{\dot{\beta}} \textrm{ , }
{F}^{\a}_{{\beta}} = \frac{i}{2} F^{\mu \nu} ({\sigma}_{\mu
 \nu})^{\a}_{{\beta}} \, \textrm{ , and } ({F}^{\a}_{{\beta}})^{\ast} = -\bar{F}^{\da}_{\dot{\beta}} \, .
\end{equation}
It satisfies the Bianchi identity
\begin{equation}
-{\cal D}_{\dot{\beta}}^{\gamma} \bar{F}^{\dot{\beta}\dot{\gamma}} = {\cal D}^{\dot{\gamma}}_\beta F^{\beta \gamma}\,.
\end{equation}
The equations of motion obtained from (\ref{lagrangian}) read:
\begin{eqnarray}
-{\cal D}_{\dot{\beta}}^{\gamma} \bar{F}^{\dot{\beta}\dot{\gamma}}
&=& {\cal D}^{\dot{\gamma}}_\beta F^{\beta \gamma} = \frac{ig}{2}
\lbrack \phi^{AB} , {\cal D}^{\gamma\dot{\gamma}} \phi_{AB}\rbrack +
2g \{
\bar{\lambda}^{\dot{\gamma}}_A , \lambda^{\gamma A} \} \nonumber\\
{\cal D}^{\alpha \da} \bar{\lambda}_{\da\, A} &=&
i\sqrt{2}g
\lbrack \phi_{AB} , \lambda^{\alpha B} \rbrack \nonumber\\
{\cal D}^{\alpha \da} \lambda^{A}_\a &=& i\sqrt{2}g
\lbrack
\phi^{AB} , \bar{\lambda}^{\da}_{B} \rbrack \nonumber\\
{\cal D}_{\alpha \da}{\cal D}^{\alpha \da} \phi_{AB} &=& \frac{1}{2
\sqrt{2}} g \epsilon_{ABCD} \{ \lambda^{\beta C},
\lambda_{\beta}^{{\cal D}} \} -\frac{1}{\sqrt{2}}g \{
\bar{\lambda}_{\dot{\beta} A}, \bar{\lambda}^{\dot{\beta}}_{B} \}
\nonumber \\
& & + g^{2} \lbrack \phi^{CD}, \lbrack \phi_{AB}, \phi_{CD} \rbrack
\rbrack \label{eqm}
\end{eqnarray}
The supersymmetry transformations that leave the action
corresponding to (\ref{lagrangian}) invariant are
\begin{eqnarray}
Q^\a_{A} \phi^{BC} &=& -i\sqrt{2} (\delta^{B}_{A} \lambda^{C \alpha} - \delta^{C}_{A} \lambda^{B \alpha})  \nonumber\\
Q^\a_{A} \lambda_{\beta}^{B} &=& -\delta^{B}_{A} F^\a_{\beta} - ig \lbrack \phi^{BC}, \phi_{CA} \rbrack \delta_{\beta}^\a  \nonumber\\
Q^\a_{A} \bar{\lambda}_{\dot{\beta} B} &=& -\sqrt{2} {\cal D}^\a_{\dot{\beta}} \phi_{BA}  \nonumber\\
Q_{\eta E} F^{\alpha \beta} &=& \sqrt{2}g \delta_{\eta}^{\beta} \lbrack \phi_{EF}, \lambda^{\alpha F} \rbrack + (\alpha \leftrightarrow \beta)  \nonumber \\
Q_{\eta E} \bar{F}^{\da \dot{\beta}} &=& -i ({\cal
D}^{\da}_{\eta} \bar{\lambda}^{\dot{\beta}}_{E} + {\cal
D}^{\dot{\beta}}_{\eta} \bar{\lambda}^{\da}_{E})  \nonumber\\
\lbrack Q^\a_{C},{\cal D}_{\dot{\beta}\beta} \rbrack &=& 2g
\delta^\a_{\beta} \bar{\lambda}_{\dot{\beta}C} \nonumber
\end{eqnarray}
and
\begin{eqnarray}
\bar{Q}^{\da\, A} \phi^{BC} &=& i \sqrt{2} \epsilon^{ABCD} \bar{\lambda}_{D}^{\da}  \nonumber\\
\bar{Q}^{\da\, A} \bar{\lambda}_{\dot{\beta}B} &=&
-\delta^{A}_{B} \bar{F}^{\da}_{\dot{\beta}}
+ i g \lbrack \phi_{BC}, \phi^{CA} \rbrack \delta_{\dot{\beta}}^{\da}  \nonumber\\
\bar{Q}^{\da\, A} {\lambda}_{{\beta}}^{B} &=& \sqrt{2} {\cal
D}^{\da}_{{\beta}}
\phi^{BA}  \nonumber\\
\bar{Q}^{\dot{\eta} E} F^{\alpha \beta} &=& -i ({\cal D}^{\dot{\eta}
\alpha} \lambda^{\beta E} +
{\cal D}^{\dot{\eta} \beta} \lambda^{\alpha E}) \nonumber\\
\bar{Q}^{\dot{\eta} E} \bar{F}^{\da \dot{\beta}} &=&
-\sqrt{2}g \epsilon^{\dot{\eta}\da} \lbrack \phi^{EF},
\bar\lambda^{\dot{\beta}}_{F} \rbrack + (\da \leftrightarrow
\dot{\beta}) \nonumber \\
\lbrack \bar{Q}^{\da\, A},{\cal D}_{\dot{\beta}\beta} \rbrack
&=& -2g \delta^{\da}_{\dot{\beta}} {\lambda}^{A}_{{\beta}} \, . \label{susy}
\end{eqnarray}
Starting from the scalar propagator
\begin{equation}
\langle  \phi^{a}_{AB}(x_{1}) \phi^{bCD}(x_{2})  \rangle = -2
\delta^{ab} (\delta^{C}_{A} \delta_{B}^{D} - \delta^{D}_{A}
\delta_{B}^{C}) \frac{1}{4 \pi^{2}} \frac{1}{x^{2}_{12}}
\label{scalarprop}
\end{equation}
and applying the supersymmetry transformations (\ref{susy}), the
fermion and gluon propagators are found to be
\begin{eqnarray}
\langle  \hat{\lambda}^{a A}(x_{1})
\hat{\bar{\lambda}}^{b}_{B}(x_{2}) \rangle &=& -4i \delta^{ab}
\delta^{A}_{B} \frac{1}{4 \pi^{2}} \frac{\hat{x}_{12}}{(x^{2}_{12})^{2}} \label{fermprop}\\
\langle  \hat{\bar{F}}^{a}(x_{1}) \hat{F}^{b}(x_{2}) \rangle &=& -32
\delta^{ab} \frac{1}{4 \pi^{2}}
\frac{\hat{x}_{12}^{2}}{(x^{2}_{12})^{3}} \label{gluonprop}\,.
\end{eqnarray}



\end{document}